\documentclass{article}\usepackage{natbib,psfig,epsfig,emulateapj}
\newcommand{\beq}{\begin{equation}}
\newcommand{\eeq}{\end{equation}}
\newcommand{\bea}{\begin{eqnarray}}
\newcommand{\eea}{\end{eqnarray}}

\newcommand{\rem}[1]{ }
\bibpunct[,]{(}{)}{;}{a}{}{,}
\begin{document}

\title{The theory of spectral evolution of the GRB prompt emission}

\author{Mikhail V. Medvedev{}\altaffilmark{1}}
\affil{
Department of Physics and Astronomy, 
University of Kansas, KS 66045} 
\altaffiltext{1}{
also at the Institute for Nuclear Fusion, RRC ``Kurchatov
Institute'', Moscow 123182, Russia}

\begin{abstract}
We develop the theory of jitter radiation from GRB shocks containing 
small-scale magnetic fields and propagating at an angle with respect 
to the line of sight. We demonstrate that the spectra vary 
considerably: the low-energy photon index, $\alpha$, ranges from 
$0$ to $-1$ as the apparent viewing angle goes from $0$ to $\pi/2$. 
Thus, we interpret the hard-to-soft evolution and the correlation of 
$\alpha$ with the photon flux observed in GRBs as a combined effect
 of temporal variation of the viewing angle and relativistic aberration 
of an individual thin, instantaneously illuminated shell. The model 
predicts that about a quarter of time-resolved spectra should have 
hard spectra, violating the synchrotron $\alpha=-2/3$ line of death. 
The model also naturally explains why the peak of the distribution of 
$\alpha$ is at $\alpha\approx-1$. The presence of a low-energy break 
in the jitter spectrum at oblique angles also explains the appearance 
of a soft X-ray component in some GRBs and a relatively small number 
of them. We emphasize that our theory is based solely on the first 
principles and contains no {\it ad hoc} (phenomenological) assumptions.
\end{abstract}


%\maketitle

\section{Introduction}

\begin{figure}
\psfig{file=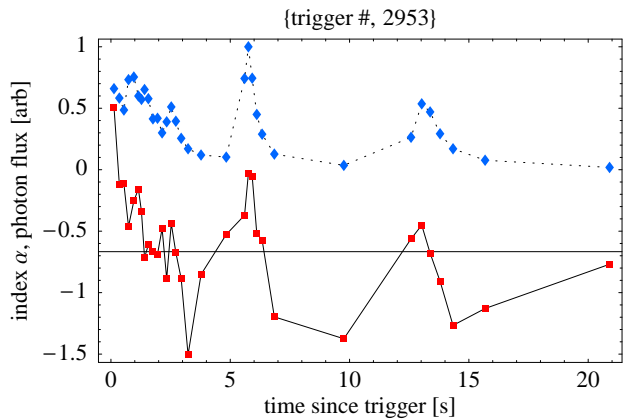,width=3.5in}
\caption{The normalized photon flux ({\it diamonds}) and the low-energy 
power-law index $\alpha$ ({\it squares}) vs time for BATSE trigger \#2953. 
The data are from the time-resolved spectral fits by \citet{Preece+00}.  
\label{f:1} }
\end{figure}

Rapid spectral variability is a remarkable, yet unexplained feature of 
the prompt GRB emission. The variation of the hardness of the spectrum 
and the hard-to-soft evolution are the most acknowledged features 
\citep{Bhat+94,Crider+97,Frontera+00,Ryde+Petrosian02}. A quite remarkable 
``tracking'' behavior, when the low-energy spectral index $\alpha$ 
follows (or correlates with) the photon flux \citep{Crider+97} is 
particularly intriguing. An example of such trend is shown in 
Fig. \ref{f:1} (we used data from \citealp{Preece+00}). 

In this paper, we demonstrate that the spectral index--flux correlation 
(or the hardness-intensity correlation) is a natural and inevitable 
prediction of the collisionless relativistic shock model and the jitter 
radiation mechanism from small-scale shock-generated magnetic fields 
\citep{ML99,M00}. Quite possibly, the appearance of a soft X-ray component 
in some GRBs \citep{Preece+95} can also be interpreted within this model. 
The theory yields a right number of the synchrotron-violating GRBs, i.e., 
with $\alpha>-2/3$, \citep{Katz94,Preece+98} and naturally explains 
why the majority of GRBs 
have $\alpha$ around $-1$ \citep{Preece+00}. We emphasize that our 
theory is based solely on the first principles and contains no 
{\it ad hoc} assumptions. 

%As a side note, we emphasize that the presented analysis confirms 
%the validity of the original results of the jitter theory \citep{M00} 
%and reinforces its applicability to GRB emission, --- the issues 
%which have recently been put into question \citep{Fleishman05}. 

\section{Theory of jitter radiation in 3D}

The angle-averaged spectral power emitted by a relativistic particle 
moving through small-scale random magnetic fields, under the assumption 
that the deflection angle is negligible and the particle trajectory 
is a straight line, has been derived elsewhere \citep{RL,LL,M00} and it reads

\beq
\frac{dW}{d\omega}=\frac{e^2\omega}{2\pi c^3}\int_{\omega/2\gamma^2}^\infty
\frac{\left|{\bf w}_{\omega'}\right|^2}{\omega'^2}
\left(1-\frac{\omega}{\omega'\gamma^2}+\frac{\omega^2}{2\omega'^2\gamma^4}
\right)\,d\omega' ,
\label{dW/dw}
\eeq
where we neglect plasma dispersion, which is vanishing for the GRB photons.
Here $\gamma$ is the Lorentz factor of a radiating particle and 
${\bf w}_{\omega'}=\int{\bf w}e^{i\omega't}\,dt$ is the Fourier  
component of the transverse particle's acceleration due to the 
Lorentz force, $F_L$. This temporal Fourier transform is taken along the 
particle trajectory, ${\bf r}={\bf r}_0+{\bf v}t$. It should 
appropriately be expressed via the statistical properties of the 
magnetic field. To make this section self-contained, we will, 
in part, follow the derivation of \citet{Fleishman05}.

We need to express the temporal Fourier component of the acceleration 
${\bf w}\equiv F_L/\gamma m$ taken along the particle trajectory in terms 
of the Fourier component of the field in the spatial and temporal domains.
Taking the Fourier transform of ${\bf w}({\bf r}_0+{\bf v}t,t)$, we have
\begin{eqnarray}
{\bf w}_{\omega'}
&=& (2\pi)^{-4}\int e^{i\omega't}\,dt 
\left( e^{-i(\Omega t-{\bf k\cdot r}_0-{\bf k\cdot v}t)} 
{\bf w}_{\Omega,{\bf k}}\, d\Omega d{\bf k} \right)
\nonumber\\
&=&(2\pi)^{-3}\int{\bf w}_{\Omega,{\bf k}}
\delta(\omega'-\Omega+{\bf k\cdot v}) \,e^{i{\bf k\cdot r}_0}
\,d\Omega d{\bf k},
\end{eqnarray}
where we used that $\int e^{i\omega t}\,dt=2\pi\delta(\omega)$. 
In a statistically homogeneous turbulence, $|{\bf w}_{\omega'}|^2$ 
should not depend on the initial point, ${\bf r}_0$, of the particle 
trajectory. Therefore, we average it over ${\bf r}_0$ as 
$\langle|{\bf w}_{\omega'}|^2\rangle
=V^{-1}\int|{\bf w}_{\omega'}|^2d{\bf r}_0$, where $V$ is the 
volume of the spatial domain. Using that 
$\int e^{i({\bf k-k}_1)\cdot{\bf r}_0}\,d{\bf r}_0
=(2\pi)^3\delta({\bf k-k}_1)$, we finally have:
\beq
\langle|{\bf w}_{\omega'}|^2\rangle
=(2\pi)^{-3}V^{-1}\int |{\bf w}_{\Omega,{\bf k}}|^2
\delta(\omega'-\Omega+{\bf k\cdot v})\,d\Omega d{\bf k}.
\label{w1}
\eeq

The Lorentz acceleration, ${\bf w}=(e/\gamma m c){\bf v\times B}$, 
can be written as 
$w_\alpha=(e/\gamma m c)\frac{1}{2}e_{\alpha \beta \gamma}
(v_\beta B_\gamma -v_\gamma B_\beta)$. Using the identity, 
$e_{\alpha \beta \gamma} e_{\alpha \lambda \mu} = \delta_{\beta \lambda} 
\delta_{\gamma \mu}-\delta_{\beta \mu} \delta_{\gamma \lambda}$, we obtain
\beq  
|{\bf w}_{\Omega,{\bf k}}|^2 
= (ev/\gamma m c)^2 (\delta_{\alpha \beta} - v^{-2} v_\alpha v_\beta)\,
B_{\Omega,{\bf k}}^\alpha B_{\Omega,{\bf k}}^{*\beta}.
\label{w2}
\eeq
In a statistically homogeneous random magnetic field, the tensor 
$B_{\Omega,{\bf k}}^\alpha B_{\Omega,{\bf k}}^{*\beta}$ can be 
expressed via the Fourier transform of the field correlation tensor
\beq
B_{\Omega,{\bf k}}^\alpha B_{\Omega,{\bf k}}^{*\beta}
= T V K_{\alpha \beta}(\Omega,{\bf k})
=TV\!\!\!\int\!\! e^{i(\Omega t-{\bf k\cdot r})} K_{\alpha \beta}({\bf r},t)
\, d{\bf r} dt,
\eeq
where $T$ is the size of the temporal domain and $K_{\alpha \beta}({\bf r},t)
= T^{-1}V^{-1}\int B_\alpha({\bf r}',t') B_\beta({\bf r'+r},t'+t)
\,d{\bf r}' dt$ is the second-order correlation tensor of the magnetic 
field \citep{Fleishman05}.

In the static case, i.e., when the magnetic field is independent of 
time, Eqs. (\ref{w1}), (\ref{w2}) read as
\begin{eqnarray}
\langle|{\bf w}_{\omega'}|^2\rangle
&=&(2\pi V)^{-1}\int|{\bf w}_{\bf k}|^2\delta(\omega'+{\bf k\cdot v})\,
d{\bf k},
\label{w1s}\\
|{\bf w}_{\bf k}|^2 
&=& (ev/\gamma m c)^2 (\delta_{\alpha \beta} - v^{-2} v_\alpha v_\beta)\,
V K_{\alpha \beta}({\bf k}).
\label{w2s}
\end{eqnarray}

\section{The magnetic field spectrum}

We adopt the following geometry: a shock is located in the $x$-$y$-plane 
and is propagating along $z$-direction. As it has initially been 
demonstrated by \citet{ML99} and later confirmed via 3D PIC simulations 
\citep{Silva+03,Nish+04,Fred+04}, the magnetic field at relativistic 
shocks is described by a random vector field in the shock plane, 
i.e., the $x$-$y$-plane. As the shock is propagating through a medium, 
the produced field is transported downstream (in the shock frame) 
whereas new field is continuously generated at the shock front. 
Thus, the field is also random in the parallel direction, i.e., the 
$z$-direction. Thus, Weibel turbulence at the shocks is 
{\em highly anisotropic}. Both the theoretical considerations and realistic 
3D simulations of relativistic shocks indicate that the dynamics of 
of the Weibel magnetic fields in the shock plane and along the normal to it
are decoupled. Hence, the Fourier spectra of the field in the $x-y$ 
plane and in $z$ direction are independent. Thus, for the Weibel fields 
at shocks, the correlation tensor has the form
\beq
K_{\alpha \beta}({\bf k})=C(\delta_{\alpha \beta}-n_\alpha n_\beta)
f_z(k_\|) f_{xy}(k_\perp),
\eeq
where ${\bf n}$ is the unit vector normal to the shock front, 
$C$ is the normalization constant proportional to $\langle B^2\rangle$, 
$f_z$ and $f_{xy}$ are the magnetic field spectra along ${\bf n}$ and in 
the shock plane, respectively, $k_\bot=(k_x^2+k_y^2)^{1/2}$ and $k_\|=k_z$, 
and finally, the tensor $(\delta_{\alpha \beta}-n_\alpha n_\beta)$ is 
symmetric and its product with ${\bf n}$ is zero, implying 
orthogonality of $\bf n$ and $\bf B$.

Numerical simulatrions \citep{Fred+04} also indicate that the field 
transverse spectrum, $f_{xy}$, is well described by a broken power-law 
with the break scale comparable to the skin depth, $c/\omega_{p}$, where 
$\omega_{p}=(4\pi e^2 n/\Gamma m)^{1/2}$ is the relativistic plasma 
frequency and $\Gamma$ is the shock Lorentz factor. We expect that the 
spectrum $f_z$, has similar properties. Therefore, we use the following models:
\beq
f_z(k_\|)=\frac{k_\|^{2\alpha_1}}{(\kappa_\|^2+k_\|^2)^{\beta_1}}, \quad 
f_{xy}(k_\bot)
=\frac{k_\perp^{2\alpha_2}}{(\kappa_\perp^2+k_\perp^2)^{\beta_2}},
\label{f}
\eeq 
where $\kappa_\|$ and $\kappa_\perp$ are parameters (being, in general, 
a function of the distance from the front, \citealp{M+05})  
determining the location 
of the peaks in the spectra, $\alpha_1,\ \alpha_2,\ \beta_1,\ \beta_2$ are 
power-law exponents below and above a spectral peak ($\beta_1>\alpha_2+1/2$ 
and $\beta_2>\alpha_2+1$, for convergence at high-$k$). Note that 
$\beta_{1,2}\to\infty$ corresponds to spectra with a sharp cut-off. 
The asymptotes of these functions are
\beq
f(k) \propto \left\{\begin{array}{ll}
k^{2\alpha}, & \textrm{if  } k \ll\kappa , \\
k^{2\alpha-2\beta}, & \textrm{if  } k\gg\kappa.
\end{array}\right.
\label{f-limits}
\eeq

\section{Radiation spectra from a shock viewed at different angles}

We now evaluate Eqs. (\ref{w1s}),(\ref{w2s}). The scalar product of 
the two tensors is
\beq
(\delta_{\alpha \beta} - v_\alpha v_\beta/v^2)
(\delta_{\alpha \beta}-n_\alpha n_\beta)=1+(n_\alpha v_\alpha)^2/v^{2}
=1+\cos^2\Theta,
\eeq
where we used that $\delta_{\alpha \alpha}=3$. Here $\Theta$ is the 
angle between the normal to the shock and the particle velocity 
(in an observer's frame), which is approximately the direction 
toward an observer, that is ${\bf v\|k}$ for an ultra-relativistic 
particle (because of relativistic beaming, the emitted radiation is 
localized within a narrow cone of angle $\sim 1/\gamma$). 
Eq. (\ref{w1s}) becomes
\beq
\langle|{\bf w}_{\omega'}|^2\rangle
=\frac{C}{2\pi}\,(1+\cos^2\Theta)\int\!\! f_z(k_\|) f_{xy}(k_\perp)
\delta(\omega'+{\bf k\cdot v})\,dk_\|d^2 k_\perp.
\label{w-main}
\eeq
Equations (\ref{dW/dw}),(\ref{w-main}) fully determine the spectrum of 
jitter radiation from a GRB shock. We now consider special cases.

\subsection{A shock viewed head-on, $\Theta=0$}

For a shock moving towards an observer, ${\bf n \| k}$, hence $\Theta=0$ 
(because $\bf n\| v$ and $\bf k\|v$ for $\gamma\gg1$) and 
${\bf k\cdot v}=k_z v$ (of course, $v\approx c$). 
Equation (\ref{w-main}) then reads
\begin{eqnarray}
\langle|{\bf w}_{\omega'}|^2\rangle
&=&\pi^{-1}C\,\overline{f_{xy}}\,\int f_z(k_z) 
\delta(\omega'+ k_z v)\,dk_z 
\nonumber\\
&=&(C/\pi|v|)\,\overline{f_{xy}}\,f_z(-\omega'/v),
\label{w-t0}
\end{eqnarray}
where $\overline{f_{xy}}=\int f_{xy}(k_\perp) d^2 k_\perp $. Apparently, 
this case is analogous to the 1D model used by \citet{M00}. We now 
determine low-$\omega$ and high-$\omega$ asymptotics for the spectrum 
given by Eq. (\ref{f}). In order to simplify the analysis, we neglect 
the small second and third terms in the brackets in Eq.(\ref{dW/dw}) 
and assume that $\alpha_1>1/2,\ \beta_1>1$. This slightly changes the 
shape of the radiation spectrum near a peak, but does not affect the 
asymptotic behavior. We have
\beq
\frac{dW}{d\omega}\propto\omega^{2\alpha_1-2\beta_1} \int_{y_0}^\infty
\frac{y^{2\alpha_1}}{(\eta_\|^2+y^2)^{\beta_1}}\,\frac{dy}{y^2},
\label{est1}
\eeq
where $y=\omega'/\omega$,\ $\eta_\|=\kappa_\|v/\omega$, and  
$y_0=1/2\gamma^2$.

At low frequencies, $\eta_\|\gg y_0$, 
(that is, $\omega\ll\kappa_\|v\gamma^2$), the right hand 
side of Eq. (\ref{est1}) can approximately be evaluated as 
$\textrm{RHS} \sim \omega^{2\alpha_1-2\beta_1} (
\int_0^{\eta_\|} y^{2\alpha_1-2}\eta_\|^{-2\beta_1}dy
+ \int_{\eta_\|}^\infty y^{2\alpha_1-2-2\beta_1}dy )
\propto \omega^{2\alpha_1-2\beta_1} \eta_\|^{2\alpha_1-1-2\beta_1}
\propto \omega^1$.
At high frequencies $\eta_\|\ll y_0$, 
(that is, $\omega\gg\kappa_\|v\gamma^2$), the right hand side 
becomes (note, $y_0=const.$) 
$\textrm{RHS} \sim \omega^{2\alpha_1-2\beta_1} 
\int_{y_0}^\infty y^{2\alpha_1-2-2\beta_1}dy
\propto \omega^{2\alpha_1-2\beta_1}$.
Combining the results, we conclude that
\beq
\left.\frac{dW}{d\omega}\right|_{\Theta=0} \propto \left\{\begin{array}{ll}
\omega^1, & \textrm{if  } \omega\ll\kappa_\|v\gamma^2 , \\
\omega^{2\alpha_1-2\beta_1}, & \textrm{if  } \omega\gg\kappa_\|v\gamma^2 .
\end{array}\right.
\label{spectr-t0}
\eeq

\subsection{A shock viewed edge-on, $\Theta=\pi/2$}

An ultra-relativistic shock moving at an angle $\sim 1/\Gamma$ with 
respect to the line of sight is seen nearly edge-on because of 
relativistic aberration. In this case, the shock is seen as if 
${\bf n \perp k}$. Therefore $\Theta=\pi/2$ and ${\bf k\cdot v}=k_x v$, 
where we assumed that an observer is located on the $x$-axis. 
Equation (\ref{w-main}) then becomes
\begin{eqnarray}
\langle|{\bf w}_{\omega'}|^2\rangle
&=&(2\pi)^{-1}C\,\overline{f_{z}}\,\int f_{xy}(k_\perp) 
\delta(\omega'+ k_x v)\,dk_x dk_y 
\nonumber\\
&=&(C/2\pi|v|)\,\overline{f_{z}}\,\int f_{xy}(\sqrt{(\omega'/v)^2+k_y^2})
\,dk_y
\nonumber\\
&\propto& (\omega')^{2\alpha_2-2\beta_2+1}\int_0^\infty
\frac{(1+y^2)^{\alpha_2}}{(\eta_*^2+1+y^2)^{\beta_2}}\,dy,
\label{w-t90}
\end{eqnarray}
where $\overline{f_{z}}=\int f_{z}(k_\|) d k_\| $. 
In the last line, we introduced $y=k_y v/\omega'$ and 
$\eta_*=\kappa_\perp v/\omega'$. The integral in Eq. (\ref{w-t90}) 
is independent of $\omega'$ when $\eta_*\ll1$, i.e., at high 
frequencies, $\omega'\gg\kappa_\perp v$. At low frequencies, 
$\eta_*\gg1$, the integral is dominated by $y\sim\eta_*$ 
(as in Eq. [\ref{est1}]), hence it is
$\propto\eta_*^{2\alpha_2-2\beta_2+1}
\propto\omega'^{-(2\alpha_2-2\beta_2+1)}$. Thus, we have
\beq
\langle|{\bf w}_{\omega'}|^2\rangle \propto \left\{\begin{array}{ll}
(\omega')^0, & \textrm{if  } \omega'\ll\kappa_\perp v , \\
(\omega')^{2\alpha_2-2\beta_2+1}, & \textrm{if  } \omega'\gg\kappa_\perp v .
\end{array}\right.
\label{est2w}
\eeq

Comparing Eqs. (\ref{est2w}) and (\ref{f-limits}), we approximate 
$\langle|{\bf w}_{\omega'}|^2\rangle$ by a function as in Eq. (\ref{f}) 
with $\alpha=0$ and $-2\beta=2\alpha_2-2\beta_2+1$. We can now find the 
asymptotes of the spectra from Eq. (\ref{dW/dw}). The analysis is 
analogous to that of Eq. (\ref{est1}) and it yields
\beq
\left.\frac{dW}{d\omega}\right|_{\Theta=\pi/2}\propto\left\{\begin{array}{ll}
\omega^0, & \textrm{if  } \omega\ll\kappa_\perp v\gamma^2 , \\
\omega^{2\alpha_2-2\beta_2+1},&\textrm{if  }\omega\gg\kappa_\perp v\gamma^2 .
\end{array}\right.
\label{spectr-t90}
\eeq
This result is analogous to that of \citet{Fleishman05}.

\subsection{A shock viewed at oblique angles, $0<\Theta<\pi/2$}

For a shock viewed at an oblique angle, 
${\bf k}={\bf \hat x}k\sin\Theta + {\bf \hat z}k\cos\Theta$, we have 
${\bf k\cdot v}=k_x v\sin\Theta+k_z v\cos\Theta$. Hence
\begin{eqnarray}
\langle|{\bf w}_{\omega'}|^2\rangle
&=&\frac{C}{2\pi}\,\int f_z(k_z)f_{x\bar y}(k_x) 
\delta(\omega'+ {\bf k\cdot v})\,dk_z dk_x 
\nonumber\\
&=&C_*\, 
\int f_z\!\left( \frac{\omega'/v}{\cos\Theta} + k_x\tan\Theta \right) 
f_{x\bar y}(k_x)\,dk_x,
\nonumber\\
&\propto& (\omega')^\zeta\int_{-\infty}^\infty
\frac{(1+y)^{2\alpha_1}}{[\eta_1^2+(1+y)^2]^{\beta_1}}
\frac{y^{2\alpha_2}\,dy}{(\eta_2^2+y^2)^{\beta_2}},
\label{w-t?}
\end{eqnarray}
where we used that $f_z(-k_z)=f_z(k_z)$ and we defined 
$C_*=C/(2\pi|v\cos\Theta|)$ and $f_{x\bar y}(k_x)
=\int f_{xy}((k_x^2+k_y^2)^{1/2}) d k_y $. One can show that this 
function is very similar to $f_{xy}(k_\perp)$ with $k_\perp$ 
replaced with $k_x$, so we use it in our analysis. In the last 
line we introduced $\zeta=2\alpha_1-2\beta_1+2\alpha_2-2\beta_2+1$,\ 
$y=k_x v \sin\Theta/\omega'$,\ $\eta_1=\kappa_\| v \cos\Theta/\omega'$,\ 
$\eta_2=\kappa_\perp v\sin\Theta/\omega'$. If $\kappa_\perp\sim\kappa_\|$ 
and $\Theta\ll1$, then $\eta_1\gg\eta_2$ and we can conclude that 
$\langle|{\bf w}_{\omega'}|^2\rangle$ has two breaks: 
$\omega'_1=\kappa_\| v \cos\Theta$ and 
$\omega'_2=\kappa_\perp v\sin\Theta$. A more detailed numerical 
analysis and the resulting radiation spectrum are discussed below.

\section{Interpretation of prompt GRB spectra}

In the standard internal shock model, each emission episode is associated 
with illumination of a thin shell, --- an internal shock and the hot and 
magnetized post-shock material. We assume that the shell is spherical 
(at least within a cone of opening angle of $\sim1/\Gamma$ around the 
line of sight) and this shell is simultaneously illuminated for a short 
period of time. The observed photon pulse is broadened because the photons 
emitted from the patches of the shell located at larger angles, 
$\vartheta$, from the line of sight arrive at progressively later 
times \citep{Piran99}. This effect naturally explains the 
fast-rise-slow-decay lightcurves of individual pulses 
\citep{Ryde+Petrosian02,Kocevski+03}. Because of relativistic 
aberration, the apparent viewing angle, $\Theta$, is greater 
than $\vartheta$ and approaches $\Theta\sim\pi/2$ (the shell is 
seen edge-on) when $\vartheta\sim1/\Gamma$. Thus, there must be a 
tight correlation between the observed spectrum and the observed 
photon flux, because they are, in essence, different manifestations 
of the same relativistic kinematics effect. 

\begin{figure}
\psfig{file=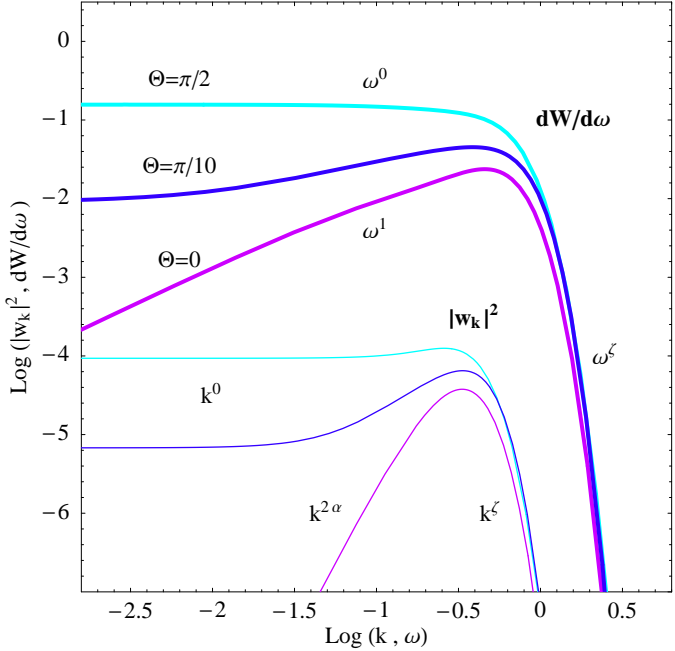,width=3.5in}
\caption{The $\log-\log$ plots of $|{\bf w_k}|^2$ vs $k$ ({\it thin lines}) 
and of $dW/d\omega$ vs $\omega$ ({\it thick lines}), for three viewing 
angles $\Theta=0,\ \pi/10,\ \pi/2$. The axes units are arbitrary. 
In this calculation we used $f_z=f_{xy}$ with 
$\alpha=2,\ \beta=20,\ \kappa=1,\ v=1$. The exponent 
$\zeta=\zeta(\alpha,\beta)$ is model dependent [c.f., 
Eqs. (\ref{spectr-t0}), (\ref{spectr-t90})]. We also chose $\gamma=1$ 
in order to align the peaks of $|{\bf w_k}|^2$ and $dW/d\omega$. 
Note that the actual peaks are at values $k,\ \omega$ lower than 
10 by a factor two or three. Note also that the spectrum $dW/d\omega$ 
levels off at oblique angles at frequencies much smaller than 
$\kappa v\gamma^2\sin\Theta$, whereas $|{\bf w_k}|^2$ indeed 
starts to flatten at $k\sim\kappa v\sin\Theta$. 
\label{f:2} }
\end{figure}

Let us now discuss specific properties of the predicted spectra. 
Fig. \ref{f:2} represents full numerical solutions of 
Eqs. (\ref{dW/dw}), (\ref{f}), (\ref{w-main}) for three different 
viewing angles. In calculation of $dW/d\omega$, the emitting electrons 
were assumed monoenergetic, for simplicity. The extension to a standard 
power-law with a sharp low-energy cutoff, $N\sim\gamma^{-p}$ for 
$\gamma>\gamma_{\rm min}$ is straightforward: the low-$\omega$ 
spectral slope remains unchanged, and the high-$\omega$ slope is 
equal to $\zeta$ or $-(p-1)/2$, whichever is greater (neglecting $e^-$ 
cooling). An important fact to note is that the jitter radiation 
spectrum varies with the viewing angle. When a shock velocity is 
along the line of sight, the low-energy spectrum is hard 
$F_\nu\propto\nu^1$, harder than the ``synchrotron line of death'' 
($F_\nu\propto\nu^{1/3}$). As the viewing angle increases, the 
spectrum softens, and when the shock velocity is orthogonal to the 
line of sight, it becomes $F_\nu\propto\nu^0$. Another interesting 
feature is that at oblique angles, the spectrum does not soften 
simultaneously at all frequencies. Instead, there appears a smooth 
spectral break, which position depends on $\Theta$. The spectrum 
approaches $\sim\nu^0$ below the break and is harder above it. 
This softening of the spectrum at low $\nu$'s could be interpreted 
as the appearance of an additional soft X-ray component, similar to 
that found in some of GRBs \citep{Preece+95}. 

\begin{figure}
\psfig{file=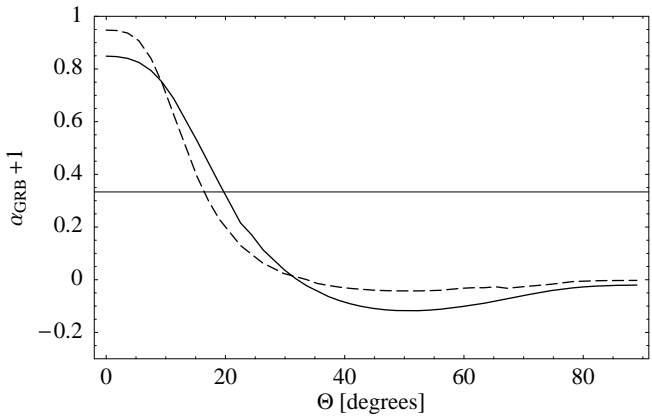,width=3.5in}
\caption{The low-energy spectral slope evaluated at $\omega\sim0.03$ 
({\em solid curve}) and $\omega\sim0.01$ ({\em dashed curve}) 
in units of Fig. \ref{f:2}, i.e., at $\omega$'s 
about 10 and 30 times below the spectral peak, corresponding to the 
low-$\omega$ edge of the {\it BATSE} window for GRBs with $E_p$ of 
about 200 and 600 keV, respectively. Note that abour 25\% of spectra 
violate the synchrotron line of death ({\it solid horizontal line}).
\label{f:3} }
\end{figure}

Fig. \ref{f:3} represents the spectral slope evaluated at frequencies 
about 10 and 30 times below the spectral peak. These frequencies 
correspond to the edge of the {\it BATSE} window for bursts with 
the peak energy of about 200~keV and 600~keV, respectively. Hence, 
the spectral slope, $\alpha_{\rm GRB}$, will be close to those 
obtained from the data fits. Since $\Theta(t)$ increases with time 
during an individual emission episode, the curves roughly represent the temporal evolution of $\alpha_{\rm GRB}$. Assuming that 
time-resolved spectra are homogeneously distributed over $\Theta$, 
one can estimate the relative fraction of the synchrotron-violating 
GRBs (i.e., those with $\alpha_{\rm GRB}+1>1/3$) as about 25\%, 
which is very close to the 30\% obtained from the data \citep{Preece+00}. 
Most of the GRBs, $\sim$75\%, should, by the same token, be distributed 
around $\alpha_{\rm GRB}\sim -1$. Note also that time-integrated GRB 
spectra should have $\alpha_{\rm GRB}$ around minus one, as well. 
This explains why a relatively large sample of synchrotron-violating 
bursts is present in the time-resolved {\it BSAX} and {\it BATSE} data
\citep{Preece+00,Frontera+00}. Since time-integrated data are dominated by 
$\alpha\sim-1$ spectra, the synchrotron-violating GRBs should be practically 
absent from the time-intergtated {\it BATSE} and {\it HETE-II} spectral 
data \citep{Barraud+03}. 
We stress that the question of why the peak of the 
$\alpha_{\rm GRB}$-distribution is at $\alpha_{\rm GRB}=-1$ and 
not at some other ``physically motivated'' value of $-3/2$ or $-2/3$, 
has had no satisfactory explanation until now. Finally, the spectral 
softening which looks like an additional soft X-ray component should 
appear within the detector spectral window when 
$\Theta\sim20^\circ\pm$~few degrees. Thus, we estimate that this 
X-ray excess can be detectable in about 10\% of GRBs, which is 
again quite close to the observed 15\% \citep{Preece+95}. 
Of course, a careful statistical analysis, which takes into 
account uneven sampling (more time-resolved spectra for brighter 
parts of the bursts), statistical and systematic errors, biases 
introduced by fits to particular spectral models, etc., is very 
desirable. However, the very fact that relative sizes of GRB 
populations fall in the right bulk part is very encouraging.

\acknowledgements

This work has been supported by NASA grant NNG-04GM41G, 
DoE grant DE-FG02-04ER54790, and the KU GRF fund.

%\clearpage
%
%

%\clearpage

%
%\clearpage

\end{document}